\documentclass[12pt,preprint]{aastex}

\slugcomment{Submitted to The Astrophysical Journal (Letters)}

\shorttitle{Low-Mass Runaways ... from Orion Trapezium}
\shortauthors{Poveda et al.}

\begin{document}

\title{Low-Mass Runaway Stars from the Orion Trapezium Cluster}


\author{Arcadio Poveda, Christine Allen and Alejandro Hern\'andez-Alc\'antara}
\affil{Instituto de Astronom\'{\i}a UNAM, Apdo. Postal 70-264,
M\'exico, D.F., 04510 M\'exico}

\begin{abstract}
In the course of a search for common proper motion binaries
in the Jones \& Walker
(JW)
catalogue of proper motions in the Orion Nebula Cluster,
we came across several faint stars with proper motions larger
than one arcsecond per century and
probabilities of membership {\it P} larger than 0.90.  Such stars
are interesting because they could be low-mass runaway stars
recently accelerated by {\it n}-body interactions in compact multiple
systems. Of particular interest among these stars is JW 451, which has
a {\it P} = 0.98, the largest transverse velocity among all the stars with
$ P \geq 0.5 $ ( $69\, \pm \, 38$  km s$^{-1}$), and a proper motion vector
which suggests that it was accelerated by the $\theta^1$ {\it
Orionis} C triple
system some 1000 years ago.  A closer examination of those JW stars
with $\mu > 1 \arcsec $ century$^{-1}$ revealed that two other stars,
JW 349 and JW 355 (with transverse velocities of $38 \pm 9 $
and $90 \pm 9 $ km s$^{-1}$ respectively), in spite of being listed
with {\it P} = 0 by JW, should also be considered part of the cluster, because
these objects are also externally ionized proplyds.  In fact, Hillenbrand
(1997) assigns to them probabilities of membership of 0.99. Moreover,
the proper motion errors of these two stars are relatively small, and so
they are good candidates to be runaway stars recently accelerated in the
Orion Nebula Cluster.
\end{abstract}


\section{Introduction}

It has been known for a long time that a subgroup of O-B stars
exhibits large peculiar velocities ($> 30$~km~s$^{-1}$). To
explain this kinematic anomaly, two theories have been advanced:
(1)  Zwicky (1957) and Blaauw (1961) proposed that when the
massive primary of a close binary explodes as a supernova of type
II, the mass ejected is sufficiently large to release the
secondary with a velocity comparable to its orbital velocity. (2)
An alternative explanation proposed by Poveda et al. (1967),
succeeded in generating large ejection velocities ($> 30$~km
s$^{-1}$) by {\it n}-body interactions from multiple stars whose initial
conditions led to very close encounters. The close interactions
frequently produced pairs of opposite runaways, like AE {\it
Auriga} and $\mu$-{\it Columbae}, the prototype of this family of
stars (Morgan \& Blaauw 1954). For a recent assessment of the two
mechanisms for producing runaways, see Hoogerwerf et al. (2001)
and references therein.

The classical runaway stars are massive O-B stars, mostly single.
However, both of the above mechanisms could also produce low mass
high-velocity stars. Is it a case of observational selection that
has made the presence of low-mass runaways to be overlooked? In a
study of internal motions of trapezium systems (Allen et al.,
1974, 2004) we pointed out the existence of some low mass stars
that appear to be leaving their host multiple systems with large
velocities.  Moreover, recent work (Loinard et al., 2003) has
revealed the existence of a low-mass star that seems to escape
from the T-{\it Tauri} system with a space velocity large enough
to qualify as a runaway.

As part of a long term investigation on the distribution of
separations of wide binaries of different ages (Poveda et al.
2004), two of the present authors (Poveda \&
Hern\'andez-Alc\'antara 2003) looked for common proper motion
binaries in the Orion Nebula Cluster, taking advantage of the very
extensive  list of precise proper motions by Jones \& Walker
(1988, henceforth JW). In the course of this investigation, our
attention was attracted by star 451 in the JW catalogue, because
its listed proper motion is the largest among all the stars with $P
>$ than 0.5, and corresponds to a relative transverse
velocity of some $69$~km s$^{-1}$ at the adopted distance of the
Orion cluster (470 pc).  The direction of the velocity vector
is such that the star seems to have been ejected from  $\theta^1$ {\it
Orionis} C.  Moreover, Jones \& Walker
assign to this star a probability of membership of $0.98$.

On closer examination, stars JW 349 and JW 355 also turn out to have
large transverse velocities ($38$~km s$^{-1}$ and $89$~km s$^{-1}$, 
respectively). Despite having {\it P} = 0 in the JW catalogue, Hillenbrand
assigns to these stars a probability of membership of 0.99 because
they have been recognized as externally ionized proplyds.

In the present letter we briefly discuss the kinematics and the nature
of these stars.  We examine whether or not they are low-mass runaway stars as
well as their probable sites of acceleration;  we propose
that their kinematic behavior could be the result of dynamical
interactions in  tight multiple systems.

\section{Jones-Walker 349}

Among the large (transverse) velocity objects in the JW catalogue we find star
349, with $\mu_{x} = $($-1.56 \pm 0.21$ )$\arcsec$ century$^{-1}$ 
and $\mu_{y} = $ ( $-0.73 \pm 0.32$ ) $\arcsec$ century$^{-1}$,
which corresponds to a transverse velocity of $38 \pm 9\, $km s$^{-1}$ assuming with JW a
distance to the Trapezium cluster of 470 pc. The large velocity and the small
errors, as well as the fact that it is an externally ionized proplyd
(Hillenbrand, 1997), make this object particularly convincing as a runaway star
in the Trapezium Cluster.

The method used by JW to determine membership to the cluster involves a
sophisticated modeling of the surface densities and motions of both cluster
members and field stars. However its application to the Trapezium Cluster is
complicated by the fact that there exists an effective absorption sheet that
practically obscures all background stars (Hillenbrand, 1997). Since all
three stars we are studying are externally ionized proplyds, one can infer
that they are indeed members of the cluster, and not projected foreground
stars.

In Table I we list the proper motions $\mu_{x}$, $\mu_{y}$ of this star together with
their one $\sigma$ errors. In Fig. 1, adapted from McCaughrean (2001), the proper
motion vectors of this star plus and minus their one $\sigma$ errors are drawn,
along with a few stars in the Orion Trapezium area. The figure suggests that
this object was ejected from the Trapezium, either from $\theta^{1}C$ or from components
$\theta^{1}A$ or $\theta^{1}B$. If it was ejected from any of these multiple systems, then it
was accelerated some 6,000 years ago. Not having a spectral type for this
object, it is difficult to estimate its mass and its age. However, given the
apparent infrared magnitude ({\it I} = 12.9) listed in JW, and the fact that it
is an externally ionized proplyd, we can infer that it is in front of the
Orion molecular cloud and embedded in the Orion H II region. Its faint
infrared magnitude is probably indicative of a low-mass object. The star
appears in JW catalogue as a variable. This is to be expected if it is in
fact an extremely young object, still in contraction towards the main sequence.

\section{Jones-Walker 355}

Star 355 in the JW catalogue is listed with a large proper motion: $\mu_{x}$ =
(3.71 $\pm$ 0.33)$\arcsec$ century$^{-1}$, $\mu_{y}$ = (-1.51 $\pm$ 0.24)$\arcsec$ century$^{-1}$
, which corresponds
to a transverse velocity of 89 $\pm$ 9 km s$^{-1}$. It has a JW probability
of membership {\it P} = 0. With this probability we should drop this star from
consideration. However, we retained it as a member of the cluster because
Hillenbrand (1997) lists it as an externally ionized proplyd with a {\it P} = 0.99,
and O'Dell \& Wong (1996) also lists it as a proplyd.  Hence its membership to
the Trapezium cluster seems reliable. Unfortunately there are no data available
on its spectrum, or photometry, apart from a JW infrared magnitude of 12.9.
The star is listed by JW as variable. The errors of the proper motion are
small, which makes the case for its large velocity quite convincing.

In Table 1 we list values of the proper motions and transverse velocities
considering errors of $\pm\sigma$. In Fig. 1 we show, superposed on the image of
the Orion Nebula Cluster, the proper motion vector of JW 355, as well as
the vectors that result from adding and subtracting to $\mu_{x}$ and $\mu_{y}$  their
one $\sigma$ errors. Projecting to the past the proper motions shown in Fig.1,
we find no object as conspicuous as  $\theta^1$ {\it
Orionis} C. However, searching within
the sector defined by the one $\sigma$ errors and up to two degrees away
($\sim$ 150,000 years back), we found an interesting object,  namely [TUK 93]28
(Tatematsu et al., 1993), a dense molecular core observed at Nobeyama in
the line CS (1-0) at 49 GHz, with an estimated mass of 260 M$_\odot$ and a radius
of $\sim$ 30,000 AU. Its line width of 1.65 km s$^{-1}$ is indicative of the presence
of young stellar objects in the core (Tatematsu et al., 1993). If the
core [TUK 93]28 is the place where JW355 was accelerated some 5,000 years
ago, it would be qualitatively similar to the scenario where the Becklin-
Neugebauer object was accelerated 500 years ago in the Kleinman-Low Complex
(Rodriguez et al., 2005). With an apparent magnitude of {\it I} = 13.6, JW 355
is clearly a low-mass star.

\section{Jones-Walker 451}
This star is listed by JW as having proper motion components:
$\mu_x $ = ($-3.06 \, \pm \, 1.49$)$\arcsec $ /century, $\mu_y $= ( $0.31\, \pm
\, 0.84$)$\arcsec$/century, with an
infrared magnitude {\it I} = 12.2 and a probability of membership {\it P} = 0.98. In
Table 1 we list the various values of the proper motion and transverse
velocities that result when considering the errors, again assuming a
distance of 470 parsecs. In Figure 1, the proper motion vector of this
star is shown. We also draw the vectors found when adding and subtracting
to $\mu_x$ and $\mu_y$ their one sigma errors ($\sigma_x$,
$\sigma_y$). The proper motion vectors of JW 451 are
consistent with the concept that this star was ejected from  $\theta^1$ {\it
Orionis} C about a thousand years ago.

The errors ($\sigma_x$,
$\sigma_y$) listed for this object are rather large, which is
understandable because of its non stellar image (proplyd). In fact, the
frequency distribution of the $\sigma_{\mu_x}$ and $\sigma_{\mu_y}$ for objects with $P > 0.5$ in
the JW catalogue, has a dispersion which is about three times larger for
the 68 proplyds than for the remaining (non-proplyd) stars.

Even if one assumes for this object the most unfavorable combination of
the components $\mu_{x}$ and $\mu_{y}$ with their errors (see vector {\it a} in Table 1 and
in Fig. 1), JW451 still remains as a runaway star coming out of the
Trapezium. In other words, with the data available, it is more likely
than not that JW451 is a runaway star ejected from the Trapezium. Clearly
this star is interesting enough to deserve further astrometric and radial
velocity studies.

TABLE 1 HERE

We have searched the literature for information on this star and have
found no additional data on its astrometry. We did find some photometric
and spectroscopic data. Hillenbrand (1997) describes this object as an
externally ionized proplyd, which confirms its membership to the Orion
Nebula Cluster and its physical proximity to  $\theta^1$ {\it
Orionis} C (see also
O'Dell \& Wen 1994); moreover, its bolometric luminosity ($L = 1.3 L_\odot$)
and its spectral type (M3e) are consistent with JW 451 being very young.
No visual magnitude for this object is available. From the values of
the $V - I$ index for the various M3e stars in Table 1 of Luhman et al.
(2000), we adopted a mean $V-I \simeq 3.3$ which, when applied to the infrared
magnitude, gives an approximate visual magnitude of  $15.5$.

With the bolometric luminosity, spectral type and temperature from Table 1
of Luhman et al. (2000), the position of JW 451 on the temperature-luminosity
diagram for the stars of the Orion Nebula cluster was plotted (see their
Fig. 6). They also plotted various gravitational contraction tracks for stars
of different masses. From this diagram, one can infer approximate values for
the mass and age of JW451. Table II summarizes the best values we have found
to characterize this object, with data taken from Hillenbrand (1997), Luhman
(2000) and Jones \& Walker (1988). The star is also listed by JW as a variable,
not surprising in view of its spectral type and its extreme youth.

TABLE 2 HERE

\section{Discussion and conclusions}

The dynamical ejection scenario turns out to be particularly intriguing when
we take into account the case of the Becklin-Neugebauer object (BN) which,
according to Tan (2004) has a transverse velocity of $38.7\,\pm\,4.7$~km s$^{-1}$
and
appears to have been ejected again from  $\theta^1$ {\it
Orionis} C  about 4000 years ago.
Where do these objects come from? Three run-away stars ejected from
$\theta^1$ {\it
Orionis}  C in six thousand years seem unlikely. Therefore we examined recent
work on the whole Becklin-Neugebauer Kleinman-Low (BN-KL) complex. A companion
paper (Rodr\'{\i}guez et al. 2005) shows convincingly
that BN was not ejected from $\theta^1$ C but rather from the
complex I-IRC$_2$, probably as the result of $n$-body
interactions.

Based on the data collected in the previous sections, we propose that JW 349
and JW451 are low-mass runaway stars that were recently ejected from the
Orion Trapezium, (see Figure 1). At present, $\theta^1$ C appears to be a hierarchical
triple system, composed of a spectroscopic binary with a period of 66 days
($a \approx 1$~AU) and a $10$ to $100$ year period speckle-resolved
companion with a mass $\geq 6 M_\odot$ (Schertl et al. 2003; Vitrichenko 2002). If JW349 or 451 were
indeed ejected from  $\theta^1$ C a few thousand years ago,  $\theta^1$ {\it
C} must have been an
unstable multiple system, at least quadruple and itself a sub-trapezium,
from which by strong dynamical interactions (given the large mass of
$\theta^1$ C) either JW451 or JW349 were ejected with a large velocity. Because of the
small mass ratio of either JW349 or 451 to  $\theta^1$ C ($\leq$ 0.01) the recoil
velocity of  $\theta^1$ C is lost within its peculiar motion; furthermore,
the present binding energy of the system  $\theta^1$ {\it
Orionis} C plus the kinetic energy
of either JW349 or JW451 remains strongly negative.

Note that components A and B of the Orion Trapezium are also known to be
multiple systems. This is particularly interesting, because high resolution
images (Figs. 4, 5 and 6 of Close et al. 2003) show  $\theta^1$ B to be a
quintuple system, and  $\theta^1$ A a triple; in both cases the similarities
of the separations
of some of the components indicate that they too are dynamically unstable
systems. These two components of the Trapezium are, in addition to
$\theta^1$ C,
good candidates for the site of acceleration of JW349 and JW451; hence,
within the uncertainties of the past trajectories of these stars, we
find three multiple stars which could have accelerated two young low-mass
runaway stars. JW355 most likely was accelerated within the dense molecular
core [TUK 93] 28.

These four runaway stars (JW349, 355, 451 and BN) must have been accelerated
by the mechanism of {\it n}-body interactions and not by any supernova explosion,
since a few thousand-year old type II supernova remnant in the Trapezium, in
the BN-KL region, or in [TUK 93] 28 would exhibit not only wild kinematics but
also the presence of very strong non-thermal synchrotron emission, neither
of which is observed.

We summarize our conclusions as follows:
(1) JW349, 355 and 451 are young, recently accelerated, runaway stars.
(2)  $\theta^1$ {\it
Orionis} A, B and C have been strongly interacting dynamically unstable
multiple systems, which appear to have produced two runaways in the last
few thousand years. (3) We have here four convincing cases (BN, JW349, JW355,
and JW 451) for runaway stars being produced by {\it n}-body interactions, and not
by a type II supernova explosion. (4) JW349, 355, 451 are good examples of
the existence of low-mass runaway stars. (5) These four runaway stars seem
to indicate that the process of star formation involves the frequent
acceleration of runaway stars in tight multiple systems.

\acknowledgments
We thank an anonymous referee for constructive suggestions which
led us to significantly alter the structure and content of this paper.

\clearpage

\begin{deluxetable}{ccccrccc}
\tablecolumns{8} \tablewidth{0pc} \tablecaption{Kinematic data for
JW 349, 355, and 451} \tablehead{ \colhead{Star}& & \multicolumn{2}{c}{$\mu_x$} &
\multicolumn{2}{c}{$\mu_y$}
& $\vert\mu\vert$ & Vt \\
& & \multicolumn{2}{c}{$''$/century}
&\multicolumn{2}{c}{$''$/century} & $''$/century & [km s$^{-1}$]}
\startdata
JW 349 & a & $+\sigma$: & -1.35 &  $-\sigma$: & -1.05 & 1.71 &  38 \\
       & b & $-\sigma$: & -1.77 &  $-\sigma$: & -1.05 & 2.06 & 46\\
       & c &            & -1.56 &             & -0.73 & 1.72 &  38\\
       & d & $+\sigma$: & -1.35 &  $+\sigma$: & -0.41 & 1.41 &  31\\
       & e & $-\sigma$: & -1.77 &  $+\sigma$: & -0.41 & 1.82 & 40\\
 & & & & & & & \\
\hline
JW 355 & a & $+\sigma$: & 4.04 &  $-\sigma$: & -1.75 & 4.40 &  98 \\
       & b & $-\sigma$: & 3.38 &  $-\sigma$: & -1.75 & 3.81 & 85\\
       & c &            & 3.71 &             & -1.51 & 4.01 &  89\\
       & d & $+\sigma$: & 4.04 &  $+\sigma$: & -1.27 & 4.23 &  94\\
       & e & $-\sigma$: & 3.38 &  $+\sigma$: & -1.27 & 3.61 & 80\\
 & & & & & & & \\
\hline
JW 451 & a & $+\sigma$: & -1.57 &  $-\sigma$: & -0.53 & 1.66 &  37 \\
       & b & $-\sigma$: & -4.55 &  $-\sigma$: & -0.53 & 4.58 & 102\\
       & c &            & -3.06 &             &   .31 & 3.08 &  69\\
       & d & $+\sigma$: & -1.57 &  $+\sigma$: &  1.15 & 1.95 &  43\\
       & e & $-\sigma$: & -4.55 &  $+\sigma$: &  1.15 & 4.69 & 104\\
\enddata
\end{deluxetable}

\clearpage

\begin{deluxetable}{ccccccc}
\tablecolumns{7} \tablewidth{0pc} \tablecaption{Parameters for
JW451} \tablehead{ \colhead{$I$} & \colhead{$V$}   &
\colhead{SP} & \colhead{$T_{{\rm eff}}$}   & \colhead{$L_{{\rm
Bol}}/L_\odot$}    & \colhead{$M/M_\odot$} & \colhead{Age
(years)}}
\startdata
12.2 & 15.5  & M3e & 3360 & 1.3 & 0.2-0.3 & 4700\\
\enddata
\end{deluxetable}

\clearpage

\begin{figure*}
\begin{center}
\includegraphics[width=12 cm,angle=0,scale=1.0]{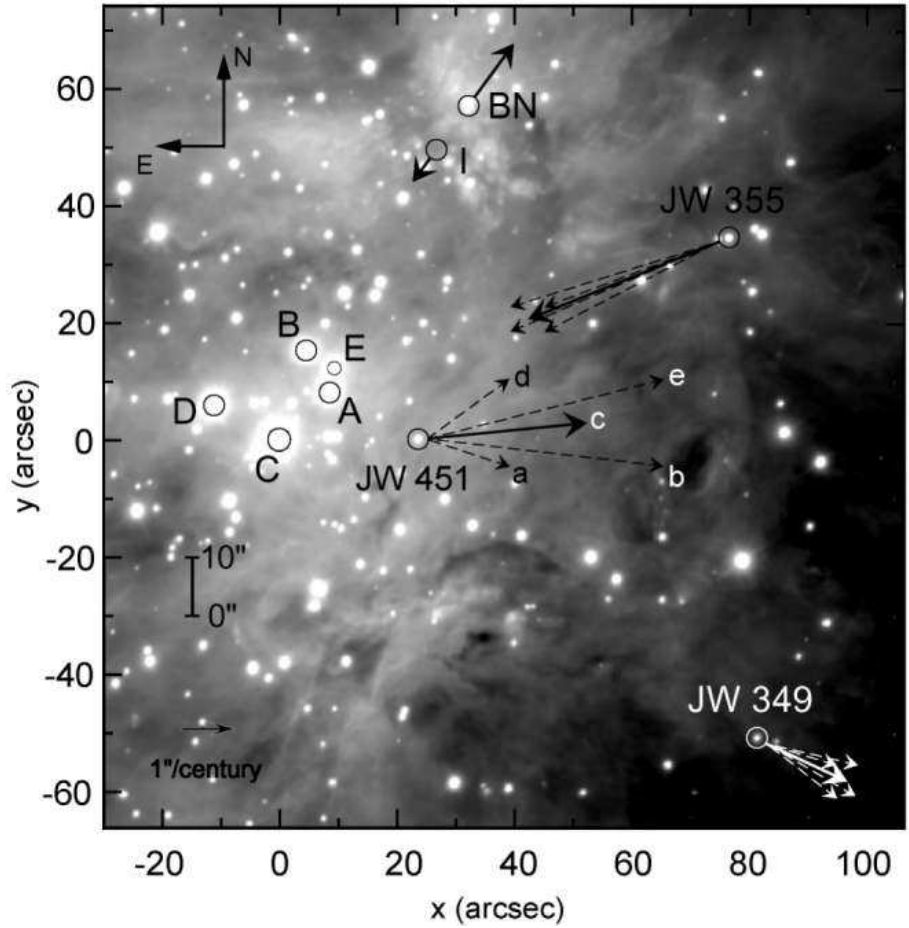}
\caption{ Proper motion vectors of runaway stars JW 349, JW 355, and 
JW 451 in relation
to the Trapezium stars, superposed on an infrared image of the
center of the Orion Nebula Cluster, from McCaughrean (2001).
Vectors a, b, c, d, e correspond to the various combinations of
$\mu$ with their $\sigma_x$, $\sigma_y$ errors.  Objects
Becklin-Neugebauer (BN) and Compact Radio Source I are also
plotted, along with their proper motion vectors (Rodriguez
et al. 2005)}
\end{center}
\end{figure*}

\end{document}